\documentclass[twocolumn,10pt]{wmrDoc}

\usepackage{epsfig} 
\usepackage{hyperref}
\hypersetup{colorlinks,linkcolor={black},citecolor={black},urlcolor={black}}  
\usepackage{graphicx} 
\usepackage[font=small,labelfont=bf]{caption}
\hypersetup{colorlinks=true,urlcolor=black}
\usepackage{tabularx}

\title{Investigating Coordinated `Social' Targeting of High-Profile Twitter Accounts}

\author{
Hunter Scott Heidenreich,
Munif Ishad Mujib, and
Jake Ryland Williams
\affiliation{
    Drexel University\\
    College of Computing and Informatics\\
    {\{hsh28, mim52, jw3477\}@drexel.edu}
}}

\begin{document}

\maketitle    

\begin{abstract}
{\it 

Following the 2016 US presidential election, 
there has been an increased focus 
on politically-motivated manipulation 
of mass-user behavior on social media platforms. 
Since a large volume of political discussion 
occurs on these platforms, 
identifying malicious activity and coordinated campaigns
is essential to ensuring 
a robust democratic environment. 
Twitter has become a critical communication channel 
for politicians and other public figures, 
enabling them to maintain 
a direct relationship with supporters. 
However, 
the platform has been fertile ground 
for large-scale malicious activity. 
As the 2020 U.S. presidential election approaches, 
we have developed tools
to monitor follower dynamics 
of some of the most prominent Twitter users, 
including U.S. presidential candidates. 
We investigate numerous, strange phenomena, 
such as dramatic spike and saw-tooth waveforms 
on follower-count charts; 
cohorts of user accounts which `circulate', 
i.e., re-follow high profile accounts numerous times; 
and other `resurrected' accounts, 
which have recently re-engaged on Twitter 
after years of non-activity. 
So through various analyses in these contexts, 
we reveal multiple, coordinated 
`social' targeting campaigns 
aimed at affecting the outcomes 
of socially critical events 
through the use of networks 
of social automations (bots),
often optimizing their social capital 
through `compromised' accounts, 
which have---unbeknownst to the greater world---been hijacked.

}
\end{abstract}

\section{Introduction}

The notion of ``social bots,"
socio-technical systems 
designed to interface with social media platforms
and interact with humans,
is not a new one.
However, as systems become more advanced,
there is increased need 
for tools that can automatically detect
these automatons
as they can have very real impacts 
on society in the form of
manipulating democracies,
inciting panic during emergencies,
and influencing major stock markets
\cite{ferrara2016rise}.

While a lone automaton might have a marginal effect
on a social media platform,
there is danger in coordinated bot accounts
(bot networks or botnets),
particularly if those networks 
are heavily embedded in public discourse.
For example, 
a covert botnet deeply embedded in public discourse
could be well positioned 
to direct and propagate misinformation
\cite{rahwan2019-etal-machine}.
In order to prevent such situations,
it becomes increasingly important 
to develop tools so that, at the very least,
humans and bots are able to recognize one another 
as they interact online.

Several works have pursued the study 
and identification of botnets on social media.
Per the taxonomy introduced in \cite{ferrara2016rise},
typical approaches follow
network-based 
\cite{stein2011facebook,cao2012aiding,paradise2014anti,xie2012innocent},
crowd sourcing \cite{wang2012social},
or feature-based \cite{ratkiewicz2011truthy,davis2016botornot,santia2019detecting} methods.
Fully coordinated botnets have previously 
been found, 
like that of the discovery 
of the `Star Wars' Twitter botnet \cite{echeverria2017discovery},
which demonstrated the existence and behavior patterns 
of systematically created bot accounts. 
Additionally, 
it is known that bot accounts 
are bought and sold in dark markets,
presenting interesting insights 
into how bot accounts occasionally trade hands 
during their lifetimes
\cite{thomas2013trafficking}. 

By monitoring statistics about users' accounts
on Twitter, 
we have identified preliminary evidence 
which appear to exhibit 
systematic and coordinated manipulation
of follower counts. 
This includes behavior that appears 
to ``buff'' (enhance) or ``nerf'' (diminish) 
the follower growth of accounts 
through superposition 
of both sharp, sub-second, delta-function spikes 
and longer ranging, sawtooth waves. 

We are presently engaged 
in investigating these phenomena further 
and the hypothesis that we are witnessing ``botnets'' 
in action---realized as weaponized 
collections of user accounts, 
with activity designed 
to either help or hurt specific individuals' statuses 
on the Twitter platform. 
Network approaches have been demonstrated 
on a case-by-case basis to be effective 
in identifying coordinated accounts in a recent study 
\cite{pacheco2020uncovering}. 
We can construct and analyze 
the relationships between groups 
of suspicious accounts 
by sampling follower lists, 
subsequently identifying discrete 
and possibly interconnected botnets. 

\begin{figure*}[th!]
    \centering
    \includegraphics[width=\textwidth]{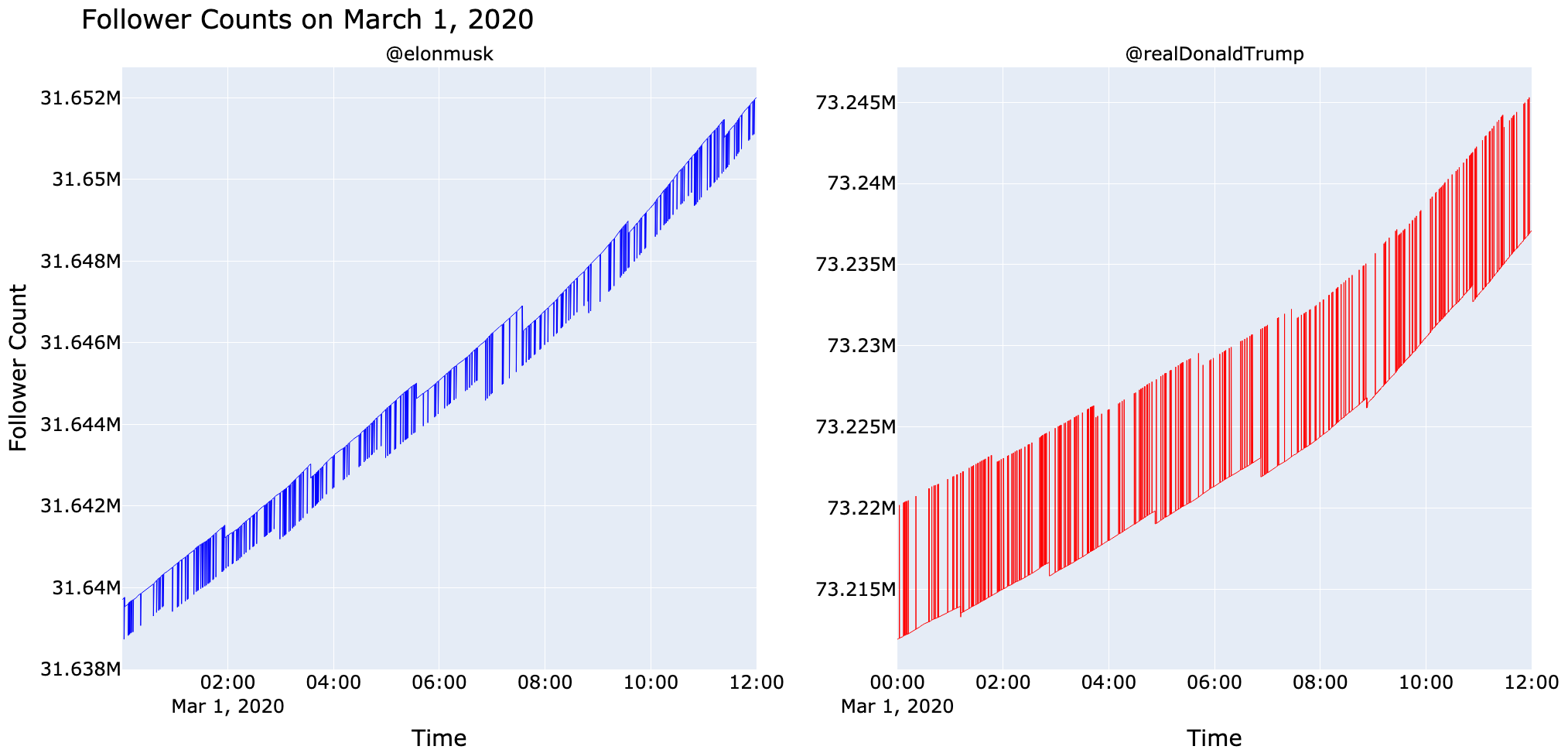}
    \caption{Follower counts of Elon Musk and Donald Trump demonstrating spike and sawtooth patterns
    over the course of 12 hours.}
    \label{fig:saws_and_spikes}
\end{figure*}

\begin{table*}[th!]
        \centering
        \caption{Observed spike and sawtooth characteristics.}
        \label{tab:spikes_and_saws}
        \scriptsize
        \resizebox{1.0\textwidth}{!}{
        \begin{tabular}{lrrrrrrr}
        \hline
        \textbf{Username} & \textbf{Followers (millions)} & \textbf{Spikes/Day} & \textbf{Avg. Spike Effect} & \textbf{Net Spike Effect} & \textbf{Sawteeth/Day} & \textbf{Avg. Sawtooth Effect} & \textbf{Net Sawtooth Effect} \\
        \hline
        \textbf{Presidential Candidates} &&&&&&& \\
        \hline
        realDonaldTrump &                 75.73 &        608 &             7,422 &      481,740,537 &           25 &               -7,907 &         -22,915,819 \\
        BernieSanders   &                 11.41 &        411 &                98 &        3,044,831 &          147 &                 -361 &          -5,699,337 \\
        JoeBiden        &                  4.70 &        361 &               -70 &       -2,511,603 &           48 &                 -123 &            -650,322 \\
        amyklobuchar    &                  1.01 &        108 &               -38 &         -392,359 &           12 &                  -28 &             -67,696 \\
        GovBillWeld     &                  0.09 &         46 &               -18 &          -43,303 &           11 &                 -113 &            -479,920 \\
        \hline
        \textbf{Individuals} &&&&&&& \\
        \hline
        elonmusk        &                 32.63 &        574 &              -924 &      -58,088,500 &           32 &               -1,053 &          -3,205,884 \\
        Cristiano       &                 83.51 &        527 &            -1,156 &      -63,668,866 &           25 &               -1,460 &          -3,229,595 \\
        ArianaGrande    &                 72.29 &        526 &              -456 &      -26,673,829 &           39 &               -1,619 &          -4,826,954 \\
        narendramodi    &                 54.63 &        506 &            -4,679 &      -95,641,313 &           26 &               -5,434 &          -8,564,685 \\
        justinbieber    &                110.71 &        492 &              -144 &       -8,839,477 &           47 &               -2,032 &          -7,782,180 \\
        \hline
        \textbf{Organizations} &&&&&&& \\
        \hline
        NASA            &                 35.91 &        505 &              -387 &      -22,288,451 &           38 &                 -806 &          -4,123,770 \\
        WhiteHouse      &                 21.12 &        502 &              -253 &      -13,426,879 &           25 &                 -512 &          -1,385,918 \\
        BBCBreaking     &                 42.37 &        493 &              -564 &      -29,563,485 &           21 &                 -861 &          -1,871,289 \\
        CNN             &                 46.43 &        459 &            -1,037 &      -16,622,269 &           42 &               -1,217 &          -2,341,954 \\
        nytimes         &                 45.83 &        449 &               -13 &        1,183,160 &           39 &                 -746 &          -3,399,156 \\
        \hline
    \end{tabular}
    }
\end{table*}

Our observations lead us 
to ask about the objectives 
of any potentially implicated botnet's actions. 
At a high level, impacts must directly target 
human perceptions of these accounts, 
or some automated aspect of Twitter's platform, 
which thus might indirectly target these perceptions.
However, given the short-range timescales 
of our observations, 
we focus on the hypothesis that 
the intended direct targets 
are Twitter's platform, 
particularly its user-search ranking utility 
as well as the recommender system 
for suggested accounts to follow.

\section{Experiments}

Largely, this work falls under 
the umbrella of the emerging sub-field 
of \emph{machine behavior} 
\cite{rahwan2019-etal-machine}. 
Operating under the hypothesis 
that these observations indicate coordinated action,
potentially due to activities of automata, 
this work describes several on-going experiments 
to better understand the phenomena observed. 
This work is primarily concerned 
with the generative mechanisms
behind theses phenomena,
and thus these experiments are designed
to serve as a basis to reason about 
such mechanisms as well as their functions.

\subsection{Coordinated (Un-)Following (E1)}

Initially, this work began in an attempt 
to measure microdynamics of Twitter-user interest 
in the political candidates participating in debate discourse.
To observe this phenomena, 
a watchlist
of 2020 U.S. Presidential election candidates'
Twitter accounts was constructed
to monitor through Twitter's API.
Having observed striking features 
in these high-resolution, temporal data 
(shown in Fig. \ref{fig:saws_and_spikes}),
the watchlist was maximally extended 
(on a free-tier set of Twitter API credentials) 
to encompass a breadth of high-profile accounts 
that are known to contribute
to political discourse 
or otherwise have grown exceptional 
social followings on the platform.

\subsubsection{Methods}

To observe changes in high profile accounts,
a cycling script was created
to monitor shifts in user profiles,
e.g., followers or friends.
Due to API rate limitations,
this script samples users 
approximately once per second.

\subsubsection{Observations}

\textbf{Mass-following/un-following events:}
Monitoring user watchlists 
exposes surprising dynamics 
on high-profile accounts. 
Fig. \ref{fig:saws_and_spikes},
exhibits follower-count data 
in raw form for two prominent users
(@realDonaldTrump and @elonmusk). 
These line charts exhibit the numbers 
of each users' followers 
at a 1-second sampling rate. 
In particular, 
Fig. \ref{fig:saws_and_spikes}
illustrates two elementary shapes:

\begin{itemize}
    \item 
    \textbf{Spikes:}
    Sharp follower count increases and decreases 
    that immediately disappear, 
    i.e. the follower count returns 
    to the pre-spike level.
    \item
    \item 
    \textbf{Sawteeth:}
    Sharp follower count decreases 
    that do not return the follower count 
    to previous levels, 
    i.e. a more permanent change compared to spikes.  
\end{itemize}

Repeated occurrences 
of both of these patterns 
strongly suggest automated behavior, 
since organic follower growth or loss 
is unlikely to occur in such a discrete manner.
For a broader cross-section 
of high profile users,
Table \ref{tab:spikes_and_saws}
characterizes the quantity,
size, and computed effect
to users grouped by broad account types.

This high sampling rate 
casts question---is it possible 
the spikes and saw-tooth features 
are algorithmic in nature? 
Or, perhaps this is a platform side-effect
caused by Twitter mass-deletions,
as has been suggested 
in the context of deleted tweets,
likes, re-tweets, and comments
\cite{minot2020ratioing}.
Unfortunately, the 1-second sampling rate 
was the limit of our standard-API capacity 
for resolution. 
All initial attempts to further resolve 
these spike and saw features 
by refining the sampling rate appeared 
to have no effect on our ability 
to observe any smoothness, 
imploring further investigation.

\subsection{Circulating Accounts (E2)}

Operating under the assumption 
that the majority of effects 
of the patterns observed 
in \textbf{E1}
are inorganic, 
one or more automated network or networks 
of Twitter accounts 
in the 10\textsuperscript{6} scale, 
at least, must exist. 
These groups of accounts must also be exhibiting 
a specific type of behavior: 
repeatedly following and un-following 
high-profile accounts. 
These \emph{circulating accounts} are key 
to understanding the phenomena of spikes and sawteeth.

\subsubsection{Methods}

A cycling download of the most recent 10,000 followers 
for each tracked account was initiated and maintained. 
API rate limitations resulted 
in an approximately 3-hour period between downloads 
for each user. 

\begin{table}[th!]
    \centering
    \caption{Observed circulation characteristics.}
    \label{tab:circs}
    \resizebox{1.0\columnwidth}{!}{
    \begin{tabularx}{0.75\textwidth}{lrrr}
        \hline
        \textbf{Username} &  \textbf{Circulating Followers} & \textbf{Circulations} & \textbf{Circulations/Day} \\
        \hline
        \textbf{Presidential Candidates} &&& \\
        \hline
         realDonaldTrump &               574,617 &      588,517 &            5,350 \\
           BernieSanders &             1,725,767 &    4,938,692 &           44,897 \\
                JoeBiden &             1,393,163 &    4,673,670 &           42,487 \\
            amyklobuchar &               239,093 &    3,917,608 &           35,614 \\
             GovBillWeld &                13,496 &    2,358,950 &           21,641 \\
        \hline
        \textbf{Individuals} &&& \\
        \hline
                elonmusk &             1,962,410 &    2,732,585 &           24,841 \\
               Cristiano &             2,460,784 &    3,905,542 &           35,504 \\
            ArianaGrande &             1,757,980 &    1,968,667 &           17,896 \\
            narendramodi &             1,381,224 &    2,653,892 &           24,126 \\
            justinbieber &             2,223,687 &    2,855,490 &           25,959 \\
        \hline
        \textbf{Organizations} &&& \\
        \hline
                    NASA &             2,263,533 &    4,423,285 &           40,211 \\
              WhiteHouse &             2,118,902 &    3,966,168 &           36,056 \\
             BBCBreaking &             2,218,925 &    4,469,857 &           40,635 \\
                     CNN &             2,136,211 &    3,233,881 &           29,398 \\
                 nytimes &             2,087,851 &    4,673,142 &           42,483 \\
        \hline
    \end{tabularx}
    }
\end{table}

\begin{figure*}[th!]
    \centering
    \includegraphics[width=\textwidth]{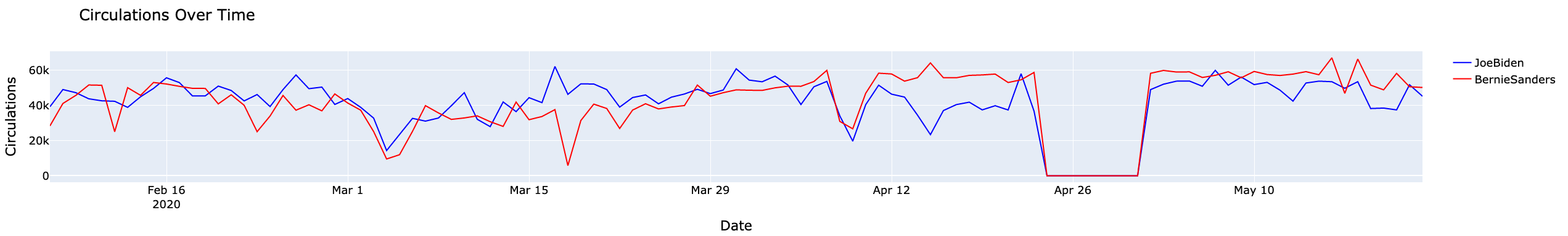}
    \caption{Circulation activity on @JoeBiden and @BernieSanders. Gap between 4/26-4/30 is due to data loss.}
    \label{fig:biden_bernie_circs}
\end{figure*}

\subsubsection{Observations}

\textbf{Limitations of Sampling:}
For many accounts, 
the number of circulating followers found 
were in the 10\textsuperscript{6} range,
as shown in Table \ref{tab:circs}. 
Critically, when follower growth rates 
are so high that significantly more followers 
than 10,000 are acquired within the 3-hour window, 
this method fails to capture the most recent, 
short-period circulation activity. 
The much lower-than-expected circulation numbers 
for @realDonaldTrump are potentially a result 
of this limitation.

\begin{figure}[bh!]
    \includegraphics[width=\columnwidth]{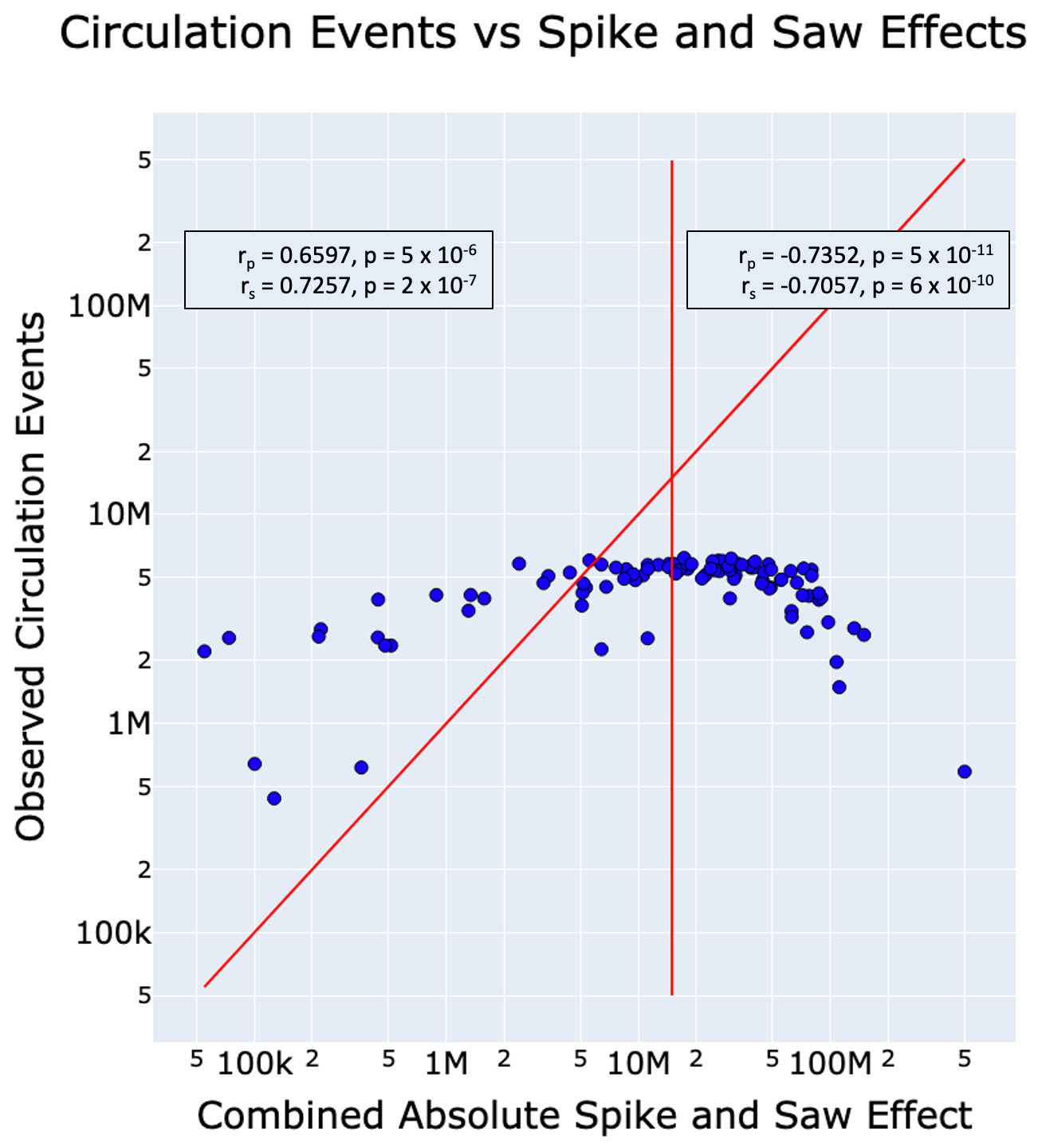}
    \caption{As the total spike and saw effect increases beyond a critical point near 15,000,000, more circulation effects presumably escape detection. 39 low-effect accounts with positive correlation and 58 high-effect accounts with negative correlation are seen.}
    \label{fig:circs_vs_spikes_and_saws}
\end{figure}

\textbf{Spikes, Saws, \& Circulations:}
Figure \ref{fig:circs_vs_spikes_and_saws} 
plots observed circulation events 
against observed spike and saw effects, 
demonstrating through correlation a seemingly 
(concave down) functional relationship between the two. 
Thus, these two entirely-separately measured effects 
are related. 
The non-monotonic behavior exhibited here 
can potentially be interpreted as follows: 
for accounts with high spike and saw effects, 
capturing all circulations becomes a challenge 
(because of rapid burying rates). 
This is evidenced by the under-estimation 
which gets `worse' as a falling trend 
in figure \ref{fig:circs_vs_spikes_and_saws}. 
At the other end, 
non-zero \href{https://developer.twitter.com/en/docs/accounts-and-users/follow-search-get-users/api-reference/get-followers-ids}{\textcolor{blue}{veracity in the ordering of follower lists}} 
(by follow time) `bakes in' some over-estimation for all,
but is only observable 
for small spike-saw effect accounts 
who generally receive too-few new followers 
to drown out this relatively marginal 
(by orders of magnitude) over-estimation. 
@realDonaldTrump is the extreme outlier at bottom right.

Figure \ref{fig:biden_bernie_circs} illustrates 
observed circulation activity 
on the final two Democratic candidates' accounts. 
Despite having half the followers of @BernieSanders,
circulation activity on @JoeBiden 
often matched or exceeded @BernieSanders'.
    
Inspecting follower lists revealed 
a further unexpected phenomenon: 
the abnormal abundance of ``ancient" Twitter accounts
among recent followers, 
especially among recent followers of @realDonaldTrump.

\subsection{Ancient Accounts (E3)}


When analyzing follower samples
of high-profile accounts
(see \textbf{E2}),
a surprising proportion 
of the samples were of older accounts.
When a Twitter users creates an account,
they are assigned a unique user id (UID). 
UIDs have a variable number of digits,
currently ranging from 2 to 19 digits,
that corresponds to when an account was created.
Older accounts have smaller UIDs (and thus fewer digits),
whereas an account created 
in July 2020 will have been assigned a 19-digit UID.

Accounts with 8-or-fewer digit UIDs correspond 
to accounts created on or before 
December 2009.
The fact that we are sampling 
follower lists of high-profile accounts---accounts
that already have significant followings---it 
is surprising that a number of ``ancient accounts"
would appear at the top of these samples,
thus indicating a recent following time.
This has lead us to question 
why these ``ancient accounts" are so active 
in recent follower samples,
and what general dynamics
Twitter's oldest users exhibit.

\subsubsection{Methods}

In order to better understand
and characterize
the behaviors and dynamics 
of ancient accounts,
a list is maintained of all
ancient accounts encountered during \textbf{E2}'s
samples.
Through another sampling script,
these ancient accounts are cycled over,
monitoring their user timelines.

\subsubsection{Observations}

\textbf{Magnitude of Occurrence}
Since beginning observation,
around \emph{2 million}
unique ancient accounts 
have appeared in the most recent 10,000 follower samples
from \textbf{E2}.
By our estimates, 
this is \emph{at least} 2\% of all of Twitter
that was created prior to December 2009,
and is around 6.5\% of the monthly active users (MAUs) 
present on the platform in 2010\footnote{
\url{https://www.statista.com/statistics/282087/number-of-monthly-active-twitter-users/},
accessed August 4th, 2020.
}.

\begin{figure*}[th!]
    \centering
    \includegraphics[width=\textwidth]{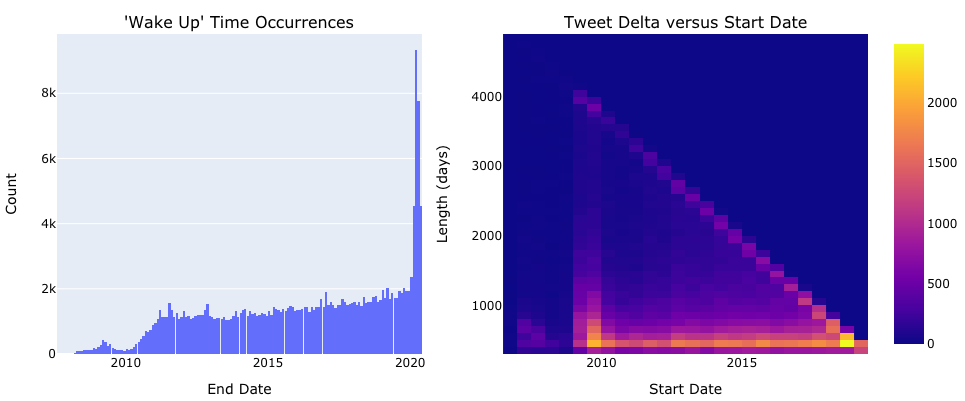}
     \caption{(Left) Histogram of end dates of tweet deltas that are longer than a year. A large number have recently started tweeting again, aligning with early US lockdown procedures and reaching a peak during 3/21-3/23. (Right) Regardless of start date, there is a strong linear correlation between length of gap and its end date due to the effects of many accounts beginning new tweet behaviors.}
    \label{fig:wakeup}
\end{figure*}

\textbf{Large Gaps in Timelines}
To better understand the nature 
of the ancient accounts observed,
timelines are currently monitored for 
all accounts with 7 or fewer digits
and all 8-digit accounts 
following @realDonaldTrump
(by-and-large one of the most dynamic
user followings, as seen in \textbf{E2}).
Of those with tweet gaps $\geq$ a year,
many appear to have recently 
``re-awoken" as seen in Fig. \ref{fig:wakeup}. 

We are compelled to ask: why did so many of these accounts recently 
end their multi-year gaps in their timelines?
Could it be a side-effect of real-world events,
perhaps of the progressing U.S. presidential primaries
or the start of COVID-19 lockdown procedures? 
Are these numbers inorganically inflated through 
activated automata?
It has previously been shown that 
some merchants that sell 
Twitter accounts create and hold these accounts 
prior to sale---sometimes upto a year even---giving 
accounts more credibility 
simply because they have been around 
on the platform longer \cite{thomas2013trafficking}.

\begin{figure}[bh!]
    \centering
    \includegraphics[width=\columnwidth]{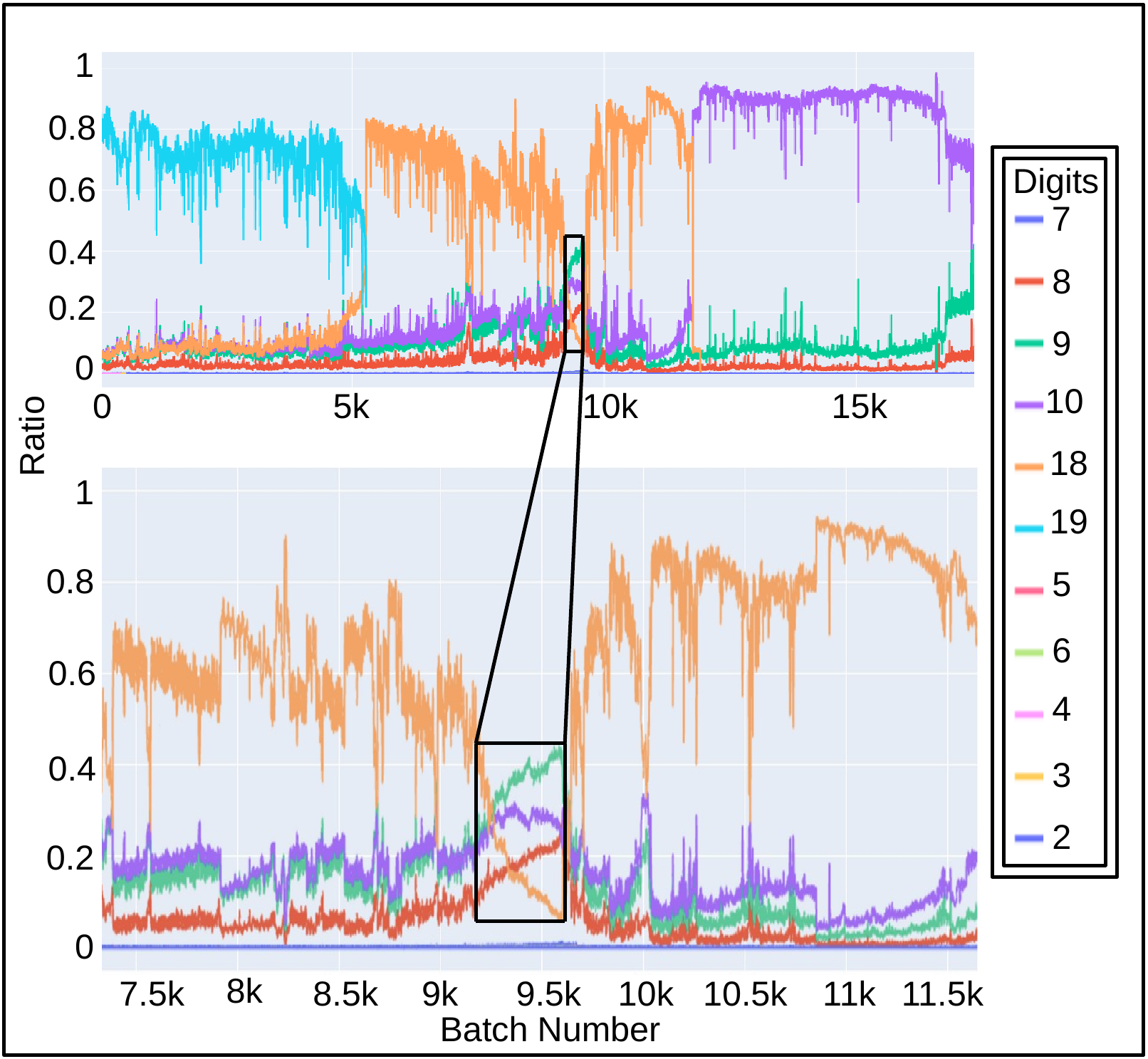}
    \caption{A fully hydrated visualization of @BarackObama's follower list, when users are grouped by digits in UID. The surprisingly inverted sub-region
    is magnified to better demonstrate 
    the inversion of digit proportions.}
    \label{fig:robot-reservoir}
\end{figure}

\subsection{High-Profile Tunneling (E4)}


In light of the surprising dynamics 
observed in follower lists
(see \textbf{E2} \& \textbf{E3}),
a series of experiments were constructed
to characterize what a full follower list 
looks like for high-profile users.

\subsubsection{Methods}

A recursive sampling script was constructed
to ``tunnel" out each follower list 
on our watchlist. 
After this process,
the UIDs were gathered 
and hydrated into full user objects
to gain more user-specific information
(e.g. privacy or verification status).

\subsubsection{Observations}

\textbf{Tunnel Dynamics:}
Full follower lists paint an interesting portrait
of the history of an account---much 
like rock strata in an excavation.
In Fig. \ref{fig:robot-reservoir}
we present what @BarackObama's follower list
looks like when followers are grouped
based on the digits in their UID. 
By grouping followers based on a given property,
one can ask questions about 
disproportionately concentrated cohorts of users.
For us, the number of digits in a UID 
coarsely approximates the account's creation date.

For example, Fig. \ref{fig:robot-reservoir}
shows an odd region of @BarackObama's following list. 
In a region dominated by 18-digit account creation,
the dynamics briefly invert deep mid-region,
with 9- and 10-digit users becoming 
the dominating groups.
This continues for about 500 batches 
of 5,000 followers (around 2.5 million followers),
before returning back to the average 
18-digit ratio between 0.6 and 0.8.

Here we are compelled to ask:
what happened to @BarackObama's followers 
that caused a mass exodus of 18-digit accounts? 
Or did an abnormally high number of older accounts
follow @BarackObama during this window? 
One could imagine scenarios where such an occurrence 
is, in fact, inorganic in nature. 
Perhaps a network of bot accounts 
embed themselves deeply on a user account's following,
for use at a later date---say, to spike 
back on the account 
or to influence long term saw-waveforms. 
So while we have not experimentally qualified this
though any temporal, following/unfollowing associations,
such a reservoir is consistent with the observation of sawteeth. 
Further experimentation is needed 
to better refine 
which hypotheses are more likely,
and whether or not this observation 
is organic or inorganic in nature.

\section{Discussion and Open Questions}

The experiments detailed in this work 
are a part of a larger set 
of diagnostic tools actively being developed 
to identify anomalous and coordinated behaviors 
on Twitter, with an emphasis 
on uncovering behaviors that may be targeting 
high-profile, and often political, users.

Many of theses behaviors are \emph{odd}.
Sub-second spikes and saw-tooth follower distortions,
disproportionate groupings of accounts 
on high-profile users,
and seemingly ancient accounts that have 
``re-awoken" after years of inactivity 
are a subset of these behaviors 
that this work highlights 
to raise awareness 
as we gather additional observations
to better disambiguate 
the \emph{how} and the \emph{why}
of the behaviors recorded.

Here, we highlight some of the open questions 
we are actively considering: 

\begin{enumerate}
    \item 
    \textbf{How can we utilize network-based methods to further pin down coordination?} 
    Previous methods have shown success in 
    both network-based and feature-based approaches,
    as highlighted above. 
    Through both friend and follower list connections,
    as well as the lesser known, general Twitter list 
    connections,
    we seek to investigate how we may apply 
    and develop new, similar methodologies 
    to highlight coordination.
    \item 
    \textbf{What is the underlying mechanism behind the behaviors observed, particularly the spike and saw-tooth behaviors?}
    Through both \textbf{E1} and \textbf{E2},
    there is strong evidence that some of this 
    behavior is a result of circulating accounts.
    However, the exact mechanism 
    as well as the underlying intent 
    of this phenomenon is still an open question.
    \item
    \textbf{How do these behaviors---particularly if inorganic---shape public discourse on Twitter and what effect do such behaviors have on international democracies?}
    Questions on how algorithms affect 
    public discourse and democracy 
    are not new \cite{ferrara2016rise,rahwan2019-etal-machine},
    and this work is very interested 
    in providing more insights into this question.
    In particular, we seek to engage in 
    semantic analyses at-scale 
    to better characterize the ``chatter" 
    suspicious accounts are engaged in,
    especially with the U.S. 2020 presidential election
    looming on the horizon.
\end{enumerate}

\section*{Acknowledgments}
This document is based upon work supported by the National Science Foundation under grant no. \#1850014.


%
% ...

\bibliographystyle{wmrDoc}



\bibliography{wmrDoc}

\end{document}